\documentclass[12pt]{article}
\usepackage{amsmath,amssymb,exscale,cancel}
\usepackage{slashed}
\usepackage{graphicx}
\usepackage{epsfig}
\usepackage{multicol}
\usepackage{color}
\usepackage{mathrsfs}
\usepackage{blindtext}
 \usepackage{fancyhdr}
\usepackage{hyperref}
\usepackage{cite}
\usepackage{mathtools}

\usepackage{floatrow}

\usepackage{graphicx}
\usepackage{sidecap}

\usepackage[latin1]{inputenc} 

\usepackage{multicol}  
 
 \usepackage{titlesec}
 
 \usepackage{rotating,slashed,xcolor,amsfonts,expdlist,charter}

\numberwithin{equation}{section}

\usepackage{xcolor}
\usepackage{sectsty}

\usepackage{mdframed}
\usepackage{titletoc}

\numberwithin{equation}{section}

\usepackage{xcolor}
\usepackage{sectsty}

\usepackage{mdframed}
\usepackage{titletoc}

\definecolor{secnum}{RGB}{13,151,225}
\definecolor{ptcbackground}{RGB}{212,237,252}
\definecolor{ptctitle}{RGB}{0,177,235}

\titlecontents{lsection}
  [5.8em]{\sffamily}
  {\color{secnum}\contentslabel{2.3em}\normalcolor}{}
  {\titlerule*[1000pc]{.}\contentspage\\\hspace*{-5.8em}\vspace*{5pt}%
    \color{white}\rule{\dimexpr\textwidth-15.5pt\relax}{1pt}}

\usepackage{hyperref}
\hypersetup{colorlinks,bookmarksopen,bookmarksnumbered,citecolor=rossos,
linkcolor=redy,pdfstartview=FitH,urlcolor=green-go}
\usepackage{slashed}

\definecolor{blus}{cmyk}{1,0.9,0,0.1}
\definecolor{verdes}{cmyk}{0.99,0,0.59,0.65}
\definecolor{rossos}{cmyk}{0,1,1,0.55}
\definecolor{redy}{cmyk}{0,1,1,0.7}
\definecolor{greeny}{cmyk}{0.99,0,0.59,0.98}
\definecolor{green-go}{cmyk}{0.79,0,0.59,0.5}

\usepackage{titlesec}

\newcommand{\beq}{\begin{equation}}
\newcommand{\eeq}{\end{equation}}

\def\hhref#1{\href{http://arxiv.org/abs/#1}{arXiv:#1}} % in bibliography

\newcommand{\tmtextbf}[1]{{\bfseries{#1}}}
\newcommand{\tmtextrm}[1]{{\rmfamily{#1}}}
\def\be{\begin{equation}}
\def\ee{\end{equation}}
\def\ba{\begin{array} }
\newcommand{\Tr}{\,{\rm Tr}}
\def\bac{\begin{array} {c}}
\def\bacc{\begin{array} {cc}}
\def\baccc{\begin{array} {ccc}}
\def\bacccc{\begin{array} {cccc}}
\def\ea{\end{array}}
\def\bea{\begin{eqnarray}}
\def\eea{\end{eqnarray}}

\definecolor{red}{rgb}{1,0,0}

\def\psl{\hbox{\hbox{${p}$}}\kern-1.9mm{\hbox{${/}$}}}
\def\dsl{\hbox{\hbox{${\partial}$}}\kern-2.2mm{\hbox{${/}$}}}
\def\Dsl{\hbox{\hbox{${D}$}}\kern-2.6mm{\hbox{${/}$}}}

\newcommand{\gappeq}{{\rlap{{\raise}.5ex\text{\ensuremath{>}}}{{\lower}.5ex\text{\ensuremath{\sim}}}}}
\newcommand{\lappeq}{{\rlap{{\raise}.5ex\text{\ensuremath{<}}}{{\lower}.5ex\text{\ensuremath{\sim}}}}}
\newcommand{\I}{\tmtextrm{1{\kern}-.24em l}}

\begin{document}
\topmargin -1.0cm
\oddsidemargin 0.9cm
\evensidemargin -0.5cm

{\vspace{-1cm}}
\begin{center}

\vspace{-1cm}

 {\Huge \tmtextbf{ 
\color{rossos} \hspace{-0.9cm}Thermal Production of Massless Dark Photons \hspace{-1.6cm}}} {\vspace{.5cm}}\\

%To do
%- submit to arXiv
%- submit to JCAP

\vspace{1.9cm}

{\large  {\bf Alberto Salvio }
{\em  
\vspace{.4cm}

 Physics Department, University of Rome Tor Vergata, \\ 
via della Ricerca Scientifica, I-00133 Rome, Italy\\

\vspace{0.6cm}

I. N. F. N. -  Rome Tor Vergata,\\
via della Ricerca Scientifica, I-00133 Rome, Italy\\

\vspace{.4cm}

%\today 

\vspace{0.4cm}

\vspace{0.2cm}

 \vspace{0.5cm}
}

\vspace{.3cm}

\vspace{0.5cm}

}
\vspace{0.cm}

\end{center}

%
% \begin{\large abstract}
% 
\noindent ---------------------------------------------------------------------------------------------------------------------------------
\begin{center}
{\bf \large Abstract}
\end{center}
\noindent A dark photon is predicted by several well-motivated Standard Model extensions and UV completions. Here  the most general effective field theory up to dimension-six operators
describing the interactions of a massless dark photon with all Standard Model particles is considered. This captures the predictions of a generic model featuring this type of vector boson at sufficiently low energies. In such framework the thermal production rate of dark photons
is computed at leading order, including the contributions of all SM particles. The corresponding cosmological yield of the dark photon  and its contribution to the effective number of neutrinos are also calculated. These predictions satisfy the current observational bounds and will be tested by future measurements.

\noindent--------------------------------------------------------------------------------------------------------------------------------

  \vspace{-0.9cm}
  
  \newpage 
\tableofcontents

\noindent

\vspace{0.5cm}

\section{Introduction}\label{Introduction}

An extra $U(1)$ gauge factor~\cite{Holdom:1985ag} with the corresponding gauge  field appear in several extensions of the Standard Model (SM)  that have been proposed to address some of the SM limitations. In order for such extra gauge field to be compatible with the experimental limits it must be somehow hidden, hence the common name ``dark photon" (DP). This is realized by requiring that all SM fields are neutral under this new $U(1)$. The physics and the observational constraints of DPs have been recently reviewed 
in~\cite{Fabbrichesi:2020wbt}.

Examples of DPs are furnished by the so called mirror world scenarios, where the observable particle physics  is duplicated and the two sectors couple to each other through gravity and perhaps other very weak forces~\cite{Berezhiani:2003xm,Berezhiani:2008gi,Salvio:2014soa}. In these constructions the DP emerges as a mirror photon.

Other examples can be found in UV completions of the SM and General Relativity. For instance, several string theory compactifications can (and generically do) lead to extra $U(1)$s, see e.g.~\cite{Abel:2003ue,Abel:2006qt,Abel:2008ai,Goodsell:2009xc}. Asymptotically safe or asymptotically free field theories can also feature DPs in their low-energy spectrum~\cite{Salvio:2020prd,Ghoshal:2020vud}.

Approaches to the cosmological constant problem~\cite{Weinberg:1988cp,Weinberg:1987dv} and/or the Higgs mass hierarchy problem~\cite{Agrawal:1997gf,DAmico:2019hih} based on the anthropic principle also motivate the presence of DPs. This is because the anthropic principle needs the construction of a multiverse, which generically have a certain number of (typically many)  DPs.

Furthermore, DPs also naturally appear in modified gravity theories where the affine connection is independent of the metric (metric-affine, Palatini and Einstein-Cartan theories), see~\cite{Baldazzi:2021kaf} for a recent overview. Indeed, in these extensions of General Relativity the affine connection can contain extra bosonic degrees of freedom of spin up to 1 in the low energy limit~\cite{Neville:1978bk,Percacci:2020ddy,Neville:1979rb,Neville:1981be,Belyaev:2007fn,Pradisi:2022nmh}.

The DP is typically accompanied by other fields, forming as a whole a ``dark sector". 
This sector is often rich enough to contain interesting features such as dark matter candidates.

 In this work the thermal production of a massless DP is computed at leading order taking into account the contribution of {\it all} SM particles and adopting a {\it model-independent} effective field theory approach. In doing so we significantly extend previous works. In Ref.~\cite{Redondo:2008ec} the thermal production of a KeV-MeV mass DP taking into account the mixing with the photon was computed.
In Ref.\cite{An:2014twa} it was pointed out that DP dark matter with a thermally generated abundance is excluded. Also, thermal production of extremely weakly coupled DPs with a mass between 1~MeV and 10~GeV was considered in~\cite{Fradette:2014sza}.  

 A massless DP can only have non-renormalizable interactions with the SM fields. Indeed, the possible kinetic mixing between the DP and the SM $U(1)_Y$ gauge field can be eliminated through field redefinitions in the massless case~\cite{Holdom:1985ag,Dobrescu:2004wz}.  Therefore, here the effective field theory needed to perform this   calculation, which features dimension-six operators, is also determined extending previous determinations~\cite{Dobrescu:2004wz}.  The higher-dimensional operators suppressed by appropriate powers of a mass scale, $M$, are interpreted as the low energy manifestation of the other fields forming the dark sector. Indeed, those extra fields can act as messengers mediating the interactions between the DP and the SM particles.
 
 This calculation of thermal DP production rate is relevant for investigating the cosmology of {\it any} model featuring a massless DP,  up to temperatures of order $M$, the cutoff of the effective field theory. Indeed, here the cosmological DP yield as a function of the temperature, $M$ and its couplings with all SM fields is calculated. Moreover, the corresponding DP contribution to the effective number of neutrinos $N_{\rm eff}^{(\nu)}$ is determined. In Ref.~\cite{Vogel:2013raa} the contribution to $N_{\rm eff}^{(\nu)}$ of massless dark photons was computed in a specific model\footnote{See also Refs.~\cite{Foot:2014uba,Adshead:2022ovo} for more recent studies.}. The computation presented here is far more general as it applies to {\it any} massless-DP model as long as one works in the regime of validity of the effective field theory. 
 
This paper is organised as follows. In Sec.~\ref{LeffSec} the relevant effective field theory including the interactions of a massless DP with all SM fields is presented. The computation of the  DP thermal production rate per unit of volume is performed in Sec.~\ref{Thermal production}, considering all possible processes at leading order. The DP yield is computed in Sec.~\ref{Dark-photon yield} and its contribution to $N_{\rm eff}^{(\nu)}$ is determined in Sec.~\ref{NeffSec}. Finally, the conclusions are provided in Sec.~\ref{Conclusions}.

\section{Effective Lagrangian}\label{LeffSec}

Let us start by presenting the most general effective Lagrangian up to dimension-six operators describing the SM fields and the gauge field $\mathcal{P}_\mu$ of an extra unbroken  Abelian group, under which all SM fields are neutral: 
\bea \mathscr{L} &=& \mathscr{L}_{\rm SM}  -\frac14 \mathcal{P}_{\mu\nu}\mathcal{P}^{\mu\nu} +\frac1{M^2}\mathcal{P}_{\mu\nu}\left(\bar Q_L \sigma^{\mu\nu} C_u \tilde H u_R+ \bar Q_L\sigma^{\mu\nu} C_d H d_R + \bar L_L \sigma^{\mu\nu} C_e H e_R +\mbox{h.c.}\right) \nonumber \\
&&
+
\frac1{M^2} \mathcal{P}_{\mu\nu}H^\dagger \left( c_b B^{\mu\nu} +\tilde c_b \tilde B^{\mu\nu}+c_w W^{\mu\nu} +\tilde c_w \tilde W^{\mu\nu}+c_p \mathcal{P}^{\mu\nu} +\tilde c_p  \mathcal{\tilde P}^{\mu\nu}\right)  H. \label{Leff}
\eea 
Here $\mathscr{L}_{\rm SM}$ represents the SM renormalizable Lagrangian and $\mathcal{P}_{\mu\nu}\equiv \partial_\mu \mathcal{P}_\nu-\partial_\nu \mathcal{P}_\mu$ is the field strength of the dark photon $\mathcal{P}_\mu$. The quantity $M$ is the mass parameter introduced in Sec.~\ref{Introduction}, which emerges by integrating out the heavy fields in the dark sector. This parameter is interpreted as the typical mass of such heavy fields, which must have some sizable couplings to the SM fields in order to generate the effective operators in~(\ref{Leff}). From the effective field theory point of view $M$ plays the role of the cutoff: the effective field theory can  only be used at energies below $M$. The first term proportional to $1/M^2$ in~(\ref{Leff}) contains all independent dimension-six operators between the DP and the SM fermions 
\be \label{fermioni} \bac Q_L\equiv \left(\bac u_L \\
d_L \ea \right)\sim (\mathbf{3}, \mathbf{2})_{1/6}\,, \quad
u_R \sim (\mathbf{3}, \mathbf{1})_{2/3}\,,\quad
d_R
\sim (\mathbf{3}, \mathbf{1})_{-1/3}\,,\\
\\
L_L\equiv \left(\bac \nu_L \\ e_L \ea \right)\sim (\mathbf{1},
\mathbf{2})_{-1/2}\,,\quad e_R \sim (\mathbf{1}, \mathbf{1})_{-1}.
 \ea \ee
$C_u$, $C_d$ and $C_e$ are generic complex $3\times 3$ matrices in flavour space, $H$ is the SM Higgs doublet
\be H \equiv \left(\bac H^+ \\ H^0 \ea \right) \sim
(\mathbf{1},\mathbf{2})_{1/2} \ee 
and
\be \tilde H\sim (\mathbf{1},\mathbf{2})_{-1/2}, \qquad \tilde H_i \equiv \epsilon_{ij} H^\dagger_j \quad (\epsilon_{ij}=-\epsilon_{ji}, \, \epsilon_{12} = 1). \ee 
 The second term proportional to $1/M^2$ in Eq.~(\ref{Leff}) describes all remaining interactions between the DP and the SM fields. They involve $H$ and the field strengths $B_{\mu\nu}$ and $W_{\mu\nu}$ of the  $U(1)_Y$ and $SU(2)_L$ SM gauge factors, respectively. For a generic field strength $\mathcal{A}_{\mu\nu}$ the dual field strength $\mathcal{\tilde A}_{\mu\nu}$ is defined  by $\mathcal{\tilde A}_{\mu\nu} \equiv \frac12\epsilon_{\mu\nu\rho\sigma} \mathcal{A}^{\rho\sigma}$, where $\epsilon_{\mu\nu\rho\sigma}$ is the totally-antisymmetric symbol with $\epsilon_{0123}=1$ (we use the mostly minus convention for the Minkowski metric $\eta_{\mu\nu}$). Finally, the parameters $c_i$ and $\tilde c_i$, with $i=b, w, p$, are real. 
 
 The operators in~(\ref{Leff}) furnish a complete basis to describe all possible interactions between the DP and the SM fields up to dimension six. Indeed, all other operators can be written as those in~(\ref{Leff}) modulo boundary terms,  operators of dimension higher than six and/or using the field equations. For example, all chirality-preserving operators even with a generic flavour structure are equivalent to the chirality-flipping ones in the first line of~(\ref{Leff}). We confirm the basis found in~\cite{Dobrescu:2004wz} with the exception of the operators with coefficients $c_w$ and $\tilde c_w$, which were missed in~\cite{Dobrescu:2004wz}. Note that there are no dimension five operators describing interactions of $\mathcal{P}_\mu$ with the SM fields only.

 \section{Thermal production  rate}\label{Thermal production}
 
 In this section we compute the thermal production rate per unit of volume of dark photons at the leading non-vanishing order, including the contribution of all SM fields in a model-independent fashion. 
 
 We assume that the temperature is above the electroweak (EW) scale such that all SM particles are in thermal equilibrium. This also allows us to neglect all masses. As will become clear in Secs.~\ref{NeffSec} and~\ref{Dark-photon yield}, this is the case in  the range of temperature that leads to the most efficient DP thermal production. The DP is very weakly coupled because it only interacts with the SM particles through dimension-six operators; therefore, the DP production can  be computed at leading order in $1/M^2$. 
 
In this section we neglect all SM masses for the reason above and use the Feynman gauge for both the SM gauge group and the dark $U(1)$. Recall that in this case $H$ contains two complex scalars.
 
 \subsection{Single dark-photon production}\label{Single dark photon production}
 
 Most of the interactions between the DP and the SM fields in~(\ref{Leff}) involve a single DP field. At leading order in $1/M^2$, the contributions of this type of interactions to the differential thermal  production rate of DPs per unit of volume can be written as follows~\cite{Bellac:2011kqa}
 \be \frac{d\gamma_d}{d^3p} = \frac{-\eta^{\mu\nu}\Pi_{\mu\nu}^<(P)}{2(2\pi)^3 P_0}, \label{DiffProd} \ee
 where $\Pi_{\mu\nu}^<(P)$ is the non time-ordered self-energy of $\mathcal{P}_\mu$ in momentum space ($P$ is its four-momentum,  $P_0=P^0$ and $\vec p$ are its energy and momentum, respectively). This quantity can be computed with the circling rules introduced by Kobes and Semenoff (KS), which generalize the cutting rules at zero temperature~\cite{Kobes:1985kc,Kobes:1986za} (see also~\cite{Bellac:2011kqa} for a textbook introduction and~\cite{Salvio:2011sf} for a summary of the KS rules in our notation). 
 %For our present purposes, all we need to know is that a circled vertex means a sign change in that vertex and a line connecting a circled vertex to an uncircled one (in the direction of the momentum) is 
 
 We now compute the various contributions to single DP production. In Sec.~\ref{Dark-photon pair production} we will compute the DP-pair production due to the operators  with coefficients $c_p$ and $\tilde c_p$ in~(\ref{Leff}), which involve two DP fields.
  
 \subsubsection{Fermion-Higgs scatterings}\label{Fermion-Higgs scatterings}
 
 We first consider the contribution of the dimension-six operators involving the SM fermions and the Higgs to the DP production (those appearing in the first line of Eq.~(\ref{Leff})).

\begin{figure}[t]
\begin{center}
  \vspace{-2.3cm}
  \includegraphics[scale=0.7]{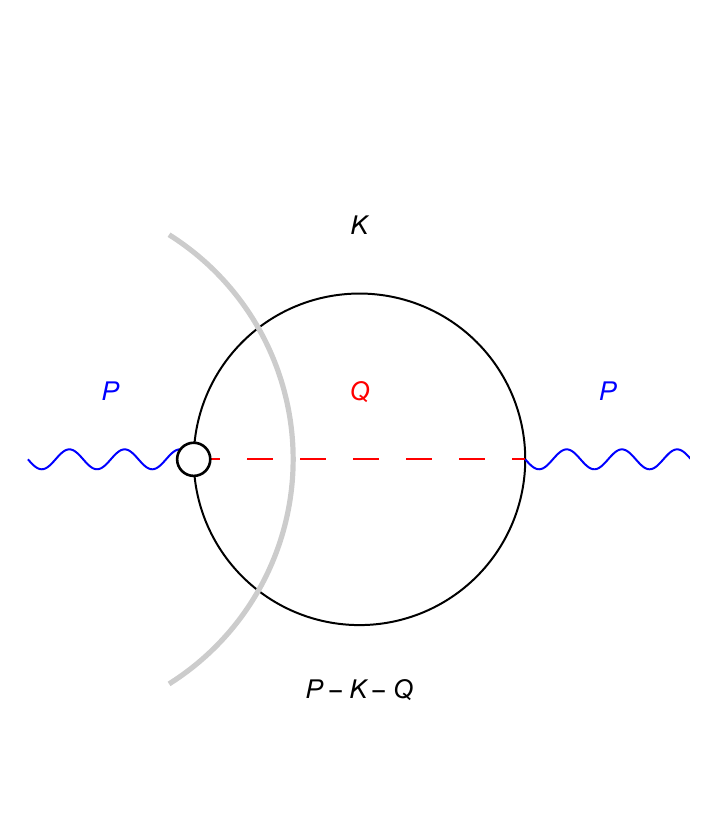}  
  \vspace{-1.2cm}
    \caption{\em The contributions to the dark photon non time-ordered self energy due to scatterings of the form~(\ref{fHsc}) involving two SM fermions (solid black lines) and a Higgs boson (dashed red line). Here, as well as in Figs.~\ref{WW} and~\ref{HHPPd}, all four-momenta flow from left to right. Here and in Figs.~\ref{WW}  we also show the corresponding cutting notation that holds at zero temperature.}
\label{fH}
  \end{center}
\end{figure}

 In this case the DP is produced through scatterings of the form 
 \be f f \to H \mathcal{P}, \qquad f H \to f \mathcal{P}, \label{fHsc}\ee 
 where the $f$s represent SM fermions. The corresponding contribution to the DP self-energy is given in Fig.~\ref{fH} and its analytic expression is 
 \be [\Pi_{\mu\nu}^<(P)]_{fH} = -\frac{4 C^2}{M^4} \int \frac{d^4K}{(2\pi)^4}\frac{d^4Q}{(2\pi)^4}\Delta_B^<(Q) \Delta_F^<(K) \Delta_F^<(P-K-Q) T_{\mu\nu}(P,K,Q), \label{SEfH}\ee
 where $C^2$ is a non-negative parameter defined in terms of the flavour matrices $C_u$, $C_d$ and $C_e$ by
 \be C^2 \equiv 2\left(3\Tr(C^\dagger_uC_u+C^\dagger_dC_d) + \Tr(C^\dagger_eC_e)\right), \label{C2def}\ee
\be  \Delta_B^<(Q) \equiv(\theta(-Q_0)+n_B(Q_0))2\pi  \delta(Q^2-m^2), \quad \Delta^<_F(K) \equiv (\theta(-K_0) - n_F(K_0))2\pi \delta(K^2-m^2) 
%= \Delta^>_F(-K) 
\nonumber\ee
($m$ is a generic mass, which we neglect here, $m\to 0$)
 and 
 \be T_{\mu\nu}(P,K,Q) \equiv P^\rho P^\lambda\left( \Tr\left(\sigma_{\rho\mu}P_R \slashed{K} \sigma_{\lambda\nu} (\slashed{K} +\slashed{Q} -\slashed{P})\right) +\Tr\left(\sigma_{\rho\mu}P_L \slashed{K} \sigma_{\lambda\nu} (\slashed{K} +\slashed{Q} -\slashed{P})\right) \right). \label{Tmunu}\ee 
 Also, $\theta(x)$ is the Heaviside step function and  $n_B(x)\equiv f_B(|x|)$, $n_F(x)\equiv f_F(|x|)$, where 
 \be f_B(x)\equiv \frac1{e^{\beta x}-1}, \quad f_F(x) \equiv \frac1{e^{\beta x} +1},\ee
 are the Bose-Einstein  and Fermi-Dirac distributions, respectively ($\beta\equiv 1/T$, where $T$ is the temperature, as usual). In~(\ref{Tmunu}) $P_R$ and $P_L$ are the right and left-handed projectors. 
 We can simply ignore $P_L$ and $P_R$ and consider only the first (or equivalently the second) trace in Eq.~(\ref{Tmunu}) because the trace of six gamma matrices times $\gamma_5$ would produce a totally-antisymmetric Levi-Civita tensor and there are not enough independent four-momenta to have a non-vanishing contraction in $\eta^{\mu\nu}\Pi^<_{\mu\nu}(P)$. By using $\sigma_{\mu\nu} = i [\gamma_\mu,\gamma_\nu]/2$ and performing the trace, we obtain
 \be -\eta^{\mu\nu} T_{\mu\nu}(P,K,Q) =  16K\cdot P(K\cdot P+Q\cdot P).  \ee
 
 The integral in~(\ref{SEfH}) receives contributions from three
distinct integration regions, which correspond to three different  scattering processes:
 \begin{enumerate}
\item $K_0 <0, \quad P_0-K_0-Q_0 >0, \quad Q_0>0 $;
\item $K_0>0, \quad  P_0-K_0-Q_0<0, \quad  Q_0>0$;
\item $K_0>0, \quad P_0-K_0-Q_0 >0,\quad  Q_0<0$.
\end{enumerate}
These regions give the following contributions to the integrated DP production rate $\gamma_d$ per unit of volume (respectively for $i=1,2,3$):
\bea [\gamma_d]_{fHi} &=& \frac{16 C^2}{(2\pi)^6 M^4} \int dp\, dk\, dq \, dz_k  \,dz_q \frac{f_i(p,k,q,z_k,z_q)}{D_i(p,k,q,z_k,z_q)^{1/2}}, \eea
where
\bea f_1(p,k,q,z_k,z_q) &=& p^2 k (1+z_k)[pk(1+z_k)-pq(1-z_q)](1-f_F(k))f_F(p+k-q) f_B(q), \nonumber  \\
f_2(p,k,q,z_k,z_q) &=& p^2 k (1-z_k)[pk(1-z_k)+pq(1-z_q)]f_F(k) (1-f_F(k+q-p))f_B(q),\nonumber\\
f_3(p,k,q,z_k,z_q) &=& p^2 k (1-z_k)[pq(1+z_q)-pk(1-z_k)]f_F(k)f_F(p-k+q)(1+ f_B(q)),\nonumber \eea
\bea D_1 (p,k,q,z_k,z_q) &\equiv& (1-z_k^2)(1-z_q^2)-\left[-1-z_k z_q+\frac{p}{q}(1+z_k)-\frac{p}{k}(1-z_q)\right]^2, \label{D1}\\
D_2 (p,k,q,z_k,z_q) &\equiv& (1-z_k^2)(1-z_q^2)-\left[1-z_k z_q-\frac{p}{q}(1-z_k)-\frac{p}{k}(1-z_q)\right]^2, \\
D_3 (p,k,q,z_k,z_q) &\equiv& (1-z_k^2)(1-z_q^2)-\left[-1-z_k z_q-\frac{p}{q}(1-z_k)+\frac{p}{k}(1+z_q)\right]^2,\label{D3}
 \eea
 $z_k$ and $z_q$ are the cosines of the angles between $\vec{k}$ and $\vec{p}$ and $\vec{q}$ and $\vec{p}$, respectively 
%\be D_1 (k,q,z_k,z_q) \equiv (1-z_k)(1-z_q)-\left[-1-z_k z_q+\frac{p}{q}(1+z_k)-\frac{p}{k}(1-z_q)\right]^2,
% \ee
 %
and the integrals are performed on the intersection between the domains 
 \begin{equation} 0\leq p < \infty, \quad 0\leq k < \infty, \quad -1 \leq z_k\leq 1, \quad 0\leq q < \infty, \quad -1 \leq z_q\leq 1 \label{IntDom0}\ee
 and
 \be  D_i (p,k,q,z_k,z_q) \geq 0. \label{IntDom}\end{equation}
 Computing numerically the integrals we obtain
 \be  [\gamma_d]_{fH} = \sum_{i=1}^3  [\gamma_d]_{fHi} \simeq 85.3 \frac{C^2 T^8}{\pi^6 M^4}. \ee

\begin{figure}[t]
\begin{center}
 \vspace{-2.3cm}
  \includegraphics[scale=0.65]{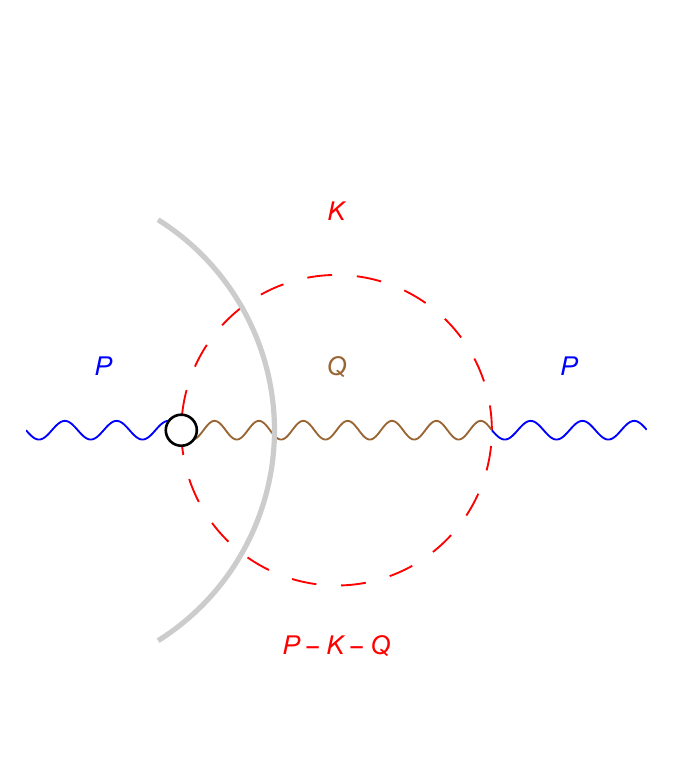} \quad  \includegraphics[scale=0.57]{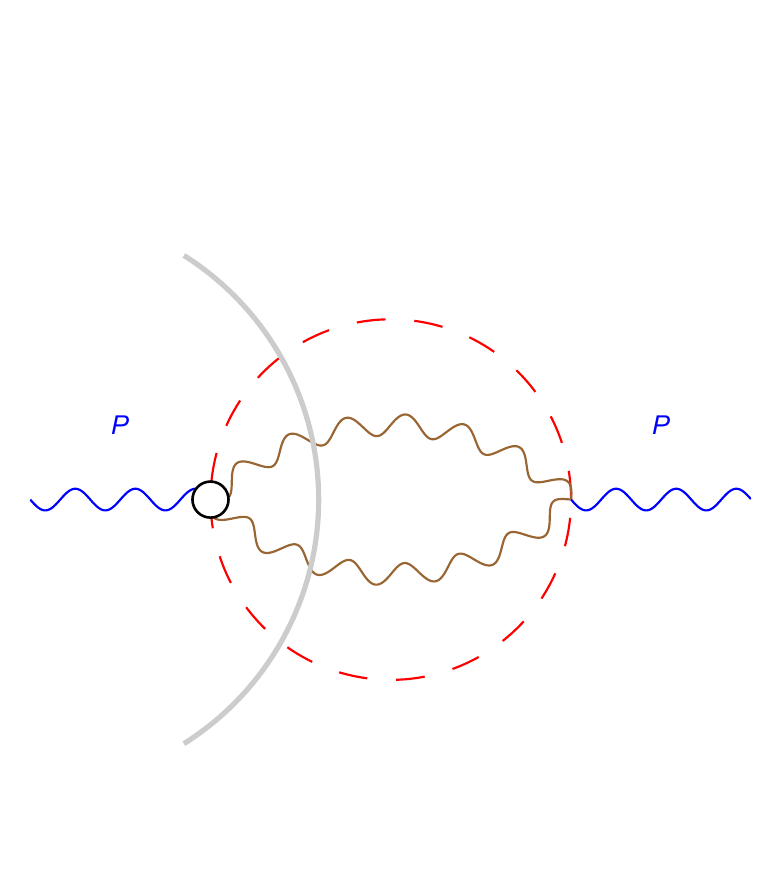}  
    \caption{\em {\bf Left:} The contributions to the dark photon non time-ordered self energy due to scatterings involving one $B$ (or one $W$) and two Higgs bosons.  {\bf Right:} The same as in the right plot, but with two (rather than one) $W$s.}
    \vspace{-1.2cm}
\label{WW}
  \end{center}
\end{figure}
 
  \subsubsection{$B$-Higgs scatterings}\label{$B$-Higgs scatterings}
  
  Let us now consider the DP production due to scattering involving one $B$ and two Higgs bosons (due to the operators with coefficients $c_b$ and $\tilde c_b$ in~(\ref{Leff})):
  \be BH \to H\mathcal{P}, \qquad HH \to B\mathcal{P}. \ee
  The corresponding contribution to the DP self-energy is given in the left plot of Fig.~\ref{WW} and its analytic expression $[\Pi_{\mu\nu}^<(P)]_{BH}$ produces the following contribution to the numerator in the right-hand side of~(\ref{DiffProd})
 \be -\eta^{\mu\nu}[\Pi_{\mu\nu}^<(P)]_{BH} = \frac{16 (c_b^2+\tilde c_b^2)}{M^4} \int \frac{d^4K}{(2\pi)^4}\frac{d^4Q}{(2\pi)^4}\Delta_B^<(Q) \Delta_B^<(K) \Delta_B^<(P-K-Q) (P\cdot Q)^2. \label{SEBH}\ee
 The interference term proportional to $c_b \tilde c_b$ vanishes because there are only three independent four-momenta, $P$, $Q$ and $K$, which are not enough to have a non-vanishing full contraction with only one Levi-Civita tensor. The contribution of $\tilde c_b^2$ is equal to that of $c_b^2$ because, with the definition of the dual field strength we are using,   
 %the dual of the dual of $B_{\mu\nu}$ is $-B_{\mu\nu}$ and
 it turns out  $B_{\mu\nu}B^{\mu\nu}=-\tilde B_{\mu\nu}\tilde B^{\mu\nu}$ and, in going from the field strength to its dual, one is only exchanging the electric and magnetic parts of the gauge field (modulo signs).

The integral in~(\ref{SEBH}) receives contributions from the  three
distinct integration regions discussed in Sec.~\ref{Fermion-Higgs scatterings}, which give the following contributions to $\gamma_d$ (respectively for $i=1,2,3$):
\bea [\gamma_d]_{BHi} &=& \frac{4 (c_b^2+\tilde c_b^2)}{(2\pi)^6 M^4} \int dp\, dk\, dq \, dz_k  \,dz_q \frac{b_i(p,k,q,z_k,z_q)}{D_i(p,k,q,z_k,z_q)^{1/2}}, \eea
where
\bea b_1(p,k,q,z_k,z_q) &=& p^3 q^2 (1-z_q)^2(1+f_B(k))f_B(p+k-q) f_B(q), \nonumber  \\
b_2(p,k,q,z_k,z_q) &=& p^3 q^2 (1-z_q)^2f_B(k) (1+f_B(k+q-p))f_B(q),\nonumber\\
b_3(p,k,q,z_k,z_q) &=& p^3 q^2 (1+z_q)^2f_B(k)f_B(p-k+q)(1+ f_B(q)),\nonumber \eea
again the functions $D_i$ are given in~(\ref{D1})-(\ref{D3}) and the integrals are performed on the intersection of the domains in~(\ref{IntDom0}) and~(\ref{IntDom}). The  numerical computation this time gives
 \be  [\gamma_d]_{BH} = \sum_{i=1}^3  [\gamma_d]_{BHi} \simeq 40.5 \frac{(c_b^2+\tilde c_b^2) T^8}{\pi^6 M^4}. \label{gammadBH}\ee

  \subsubsection{$W$-Higgs scatterings}
 
 The DP can also be produced in scatterings involving Higgs and $W$ bosons. In this case there are two contributions to the non time-ordered DP self energy. 
 
 One is analogous to that corresponding to scatterings involving the Higgs and the $B$ bosons, which have been discussed in Sec.~\ref{$B$-Higgs scatterings}. This contribution to $\gamma_d$ can be obtained by substituting $(c_b, \tilde c_b) \to  (c_w, \tilde c_w)$ in~(\ref{gammadBH}):
  \be  [\gamma_d]_{WH}  \simeq 40.5 \frac{(c_w^2+\tilde c_w^2) T^8}{\pi^6 M^4}. \label{gammadWH}\ee

 The other contribution is due to the non-Abelian nature of $W$, which leads to a term proportional to $[W_\mu,W_\nu]$ in $W_{\mu\nu}$ and diagrams of the form given in Fig.~\ref{WW} (on the right).  This  leads to a vanishing contribution to the differential production rate $\frac{d\gamma_d}{d^3p}$. Indeed, the tensorial structure of the corresponding contribution to the non time-ordered DP self energy, $[\Pi^<_{\mu\nu}(P)]_{WW}$, is proportional to $P_\mu P_\nu$ (as in both vertices there is a derivative acting on the massless DP field and no derivatives on the other fields\footnote{Recall that we use the Feynman gauge so the tensorial structure of the gauge field propagators is proportional to $\eta_{\mu\nu}$.}) and so, since the DP is massless, $\eta^{\mu\nu}[\Pi^<_{\mu\nu}(P)]_{WW}=0$.

 \subsection{Dark-photon pair production}\label{Dark-photon pair production}
 
 \begin{figure}[t]
\begin{center}
 \vspace{-2.9cm}
  \includegraphics[scale=0.6]{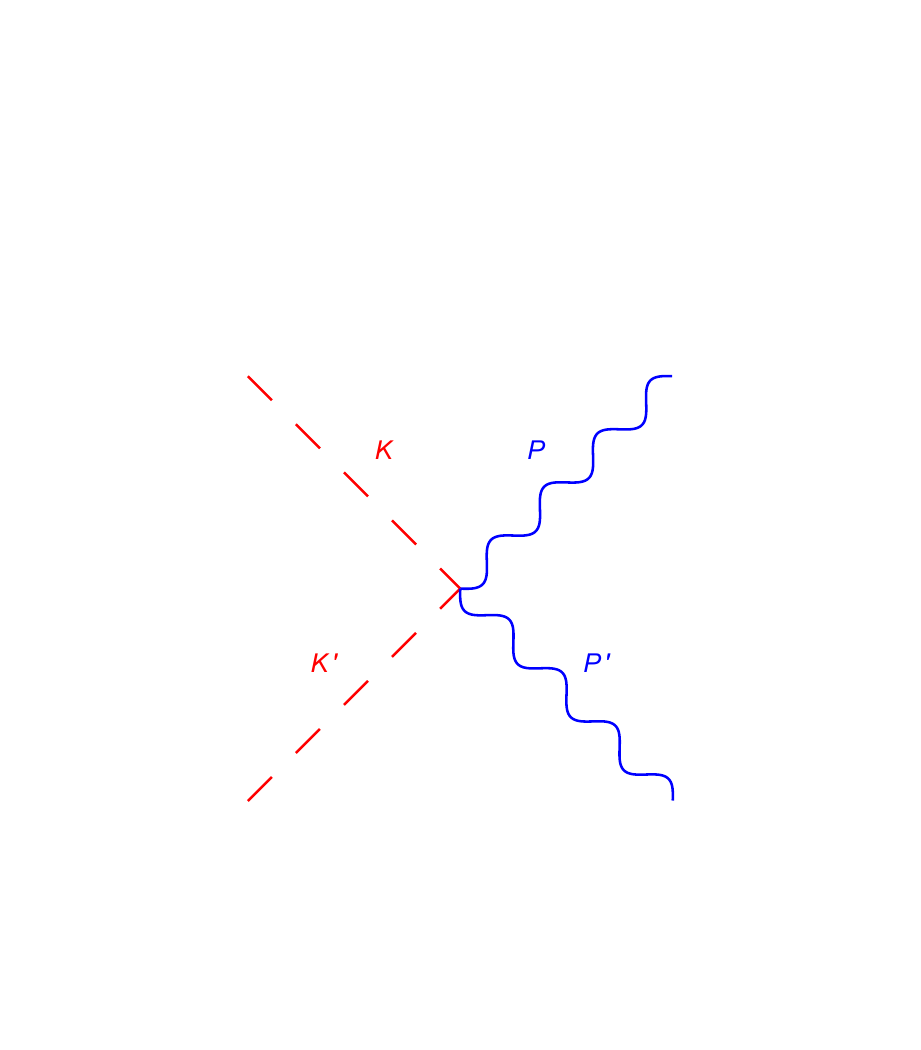}  
  \vspace{-1.8cm}
    \caption{\em Dark-photon pair production due to Higgs pair annihilation.}
\label{HHPPd}
  \end{center}
\end{figure}

 Let us now turn to the contribution of the operators  with coefficients $c_p$ and $\tilde c_p$ in~(\ref{Leff}). These lead to DP pair production through scatterings of the form (Higgs pair annihilation)
 \be H H \to \mathcal{P} \mathcal{P}, \label{HHPP}\ee 
 whose Feynman diagram is given in Fig.~\ref{HHPPd}.

 The Feynman amplitude corresponding to Fig.~\ref{HHPPd}  is  
\be \mathcal{M}(H H \to \mathcal{P} \mathcal{P})=\frac{4 i }{M^2}\left[c_p (P\cdot \varepsilon' \,  P'\cdot\varepsilon- P\cdot P' \, \varepsilon\cdot\varepsilon')+\tilde c_p P_\mu P'_\sigma \varepsilon_\nu\varepsilon'_\rho\epsilon^{\mu\nu\rho\sigma}\right], \ee
 where $\varepsilon$ and $\varepsilon'$ are the polarization vectors of the DPs with four-momenta $P$ and $P'$, respectively. Summing over the respective polarizations $r$ and $r'$ and multiplying by a factor of 2 to take into account that the Higgs is a doublet, one finds
 \be 2\sum_{rr'} |\mathcal{M}(H H \to \mathcal{P} \mathcal{P})|^2= \frac{64 (c_p^2+\tilde c_p^2) (P\cdot P')^2}{M^4}.\ee
 
 Again, the interference term (in this case proportional to $c_p\tilde c_p$)  vanishes: after summing over the polarizations, all Lorentz indices of the four-momenta and of only one  Levi-Civita tensor  are contracted together and there are not enough independent four-momenta to have a non-vanishing result; also, the contribution of $\tilde c_p^2$ is equal to that of $c_p^2$ because  
 % the dual of the dual of $\mathcal{P}_{\mu\nu}$ is $-\mathcal{P}_{\mu\nu}$ and
  $\mathcal{P}_{\mu\nu}\mathcal{P}^{\mu\nu}=-\mathcal{\tilde P}_{\mu\nu}\mathcal{\tilde P}^{\mu\nu}$.

%The interference term proportional to $c_b \tilde c_b$ vanishes because there are only three independent momenta $P$, $Q$ and $K$ which is not enough to have a non-vanishing contraction with the Levi-Civita tensor.

 The corresponding DP-pair production rate per unit of volume and averaged over the initial state with the Bose-Einstein distributions of the two Higgs particles is then
 \be \gamma_{2d} = \int\frac{d^3k}{2(2\pi)^3 E}f_B(E)\frac{d^3k'}{2(2\pi)^3 E'} f_B(E') \, I(Q^2),  \ee
 where $E$ and $E'$ are the energies of the Higgs bosons with four-momenta $K$ and $K'$ respectively and $I$ is a Lorentz-invariant function of $Q\equiv K+K'$ given by
 \be I(Q^2)= \int\frac{d^3p}{2(2\pi)^3 E_p}\frac{d^3p'}{2(2\pi)^3 E'_p} (2\pi)^4\delta(P+P'-Q)\, 2\sum_{rr'} |\mathcal{M}(H H \to \mathcal{P} \mathcal{P})|^2 \ee
 ($E_p$ and $E_p'$ are the energies of the two DPs). By performing the integrals over $\vec p$ and $\vec p\, '$ one obtains 
 \be I(Q^2)  = \frac{c_p^2+\tilde c_p^2}{\pi} \left(\frac{Q^2}{M^2}\right)^2 \ee
% \xxx{I divided by a factor of two to take into account that the two DPs in the final state are identical. This change propagates to Eq.~(\ref{g2d}) and~(\ref{alpha})}
 and, integrating over $\vec k$ and $\vec k\,'$,
 \be \gamma_{2d} =  \frac{(c_p^2+\tilde c_p^2)\pi^3 T^8}{675 M^4}. \label{g2d}\ee
 
  \subsection*{Summary}
  The full rate of DP thermal production per unit of volume is obtained by summing the contributions obtained in the previous sections~\ref{Single dark photon production} and~\ref{Dark-photon pair production}:
  \be \gamma_d = \frac{\alpha T^8}{\pi^3 M^4},  \label{gammadSum}\ee
where
\be \alpha \simeq \left[2.75\,  C^2+1.31 (c_b^2+\tilde c_b^2+c_w^2+\tilde c_w^2)+ \frac{2\pi^6(c_p^2+\tilde c_p^2)}{675}\right] \label{alpha} \ee
and  $C^2$ is defined in Eq.~(\ref{C2def}).

 \section{Dark-photon yield}\label{Dark-photon yield}
  Let us now compute the DP yield as a function of the temperature, which is important for determining the DP abundance during the cosmological history. 
  %The low-energy effective field theory~(\ref{Leff}) does not contain a DM candidate and a mechanism for neutrino masses and baryogenesis. Therefore, we assume inflation and reheating have taken place before the temperature dropped below $M$ such that a DM candidate and a suitable mechanism for neutrino masses and baryogenesis is provided by the dark sector, which features, as discussed, heavy fields with mass $\sim M$. 
  
After inflation and reheating have taken place, the relevant Boltzmann equation  is
 \be sH T \frac{dY_d}{dT} =-\gamma_d \left(1-\frac{Y_d}{Y_d^{\rm eq}} \right). \label{BolEq} \ee
The various quantities that appear in this equation are defined as follows.
 \begin{itemize}
 \item $s$ is the entropy density of the relativistic SM plasma $s= 2  \pi^2  g_{\rm SM} T^3/45$, with $g_{\rm SM}\simeq 427/4$.
 %all SM particles are relativistic otherwise the calculation of $\gamma_d$ above is not valid
 %there may be an extra +2 due to the DP if its presence is statistically relevant in the universe.
\item 
 $H$ is the Hubble rate $H = \sqrt{\rho/3}/M_P$, where  $M_P$ is the reduced Planck mass and $\rho \simeq \pi^2 g_{\rm SM}T^4/30$.
 %there may be an extra +2 to add to $g_{\rm SM}$  due to the DP if its presence is statistically relevant in the universe
 \item The DP yield $Y_d$ is defined as the comoving number density, $Y_d\equiv n_d/s$, where $n_d$ is the DP number density with equilibrium value $n_d^{\rm eq} = 2\zeta(3)T^3/\pi^2$ and $\zeta(s)$ is the Riemann zeta function.  %eq.(5.4) of Fabbrichesi - Gabrielli - Lanfranchi  2005.01515
So $Y_d^{\rm eq} =  n_d^{\rm eq}/s\simeq 0.005$. 
 \end{itemize}
 
  \begin{figure}[t]
\begin{center}
  \includegraphics[scale=0.7]{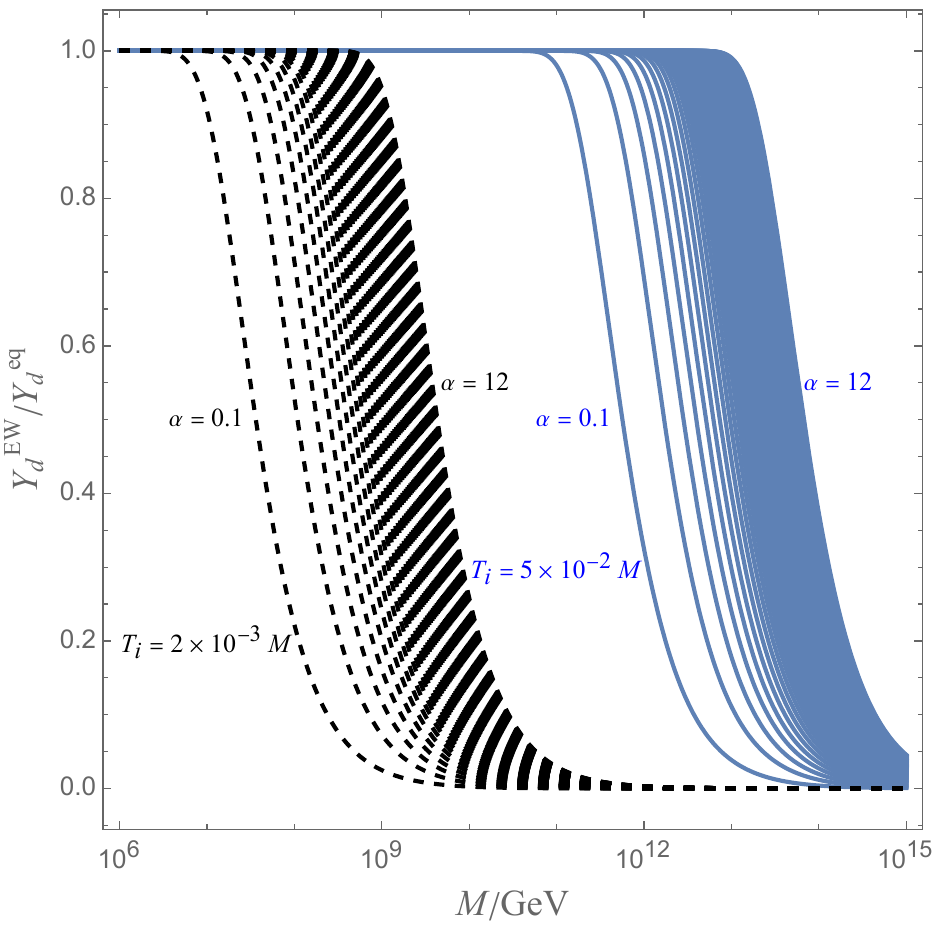}  
  \vspace{-0cm}
    \caption{\em The dark-photon yield at the EW scale $T\sim 250$~GeV as a function of $M$ setting $Y_d(T_i) = 0$ at the initial temperature $T_i$.  The parameter $\alpha$ is defined in Eq.~(\ref{alpha}) and is increased in the plot with steps of length $0.2$. }
\label{YvM}
  \end{center}
\end{figure}

 Using~(\ref{gammadSum}), the general solution of Eq.~(\ref{BolEq}) is 
 \be Y_d(T) = Y_d^{\rm eq} + (Y_i-Y_d^{\rm eq})\exp\left(\frac{45}{\pi^6}\sqrt{\frac{5}{2g_{\rm SM}^3}} \frac{\alpha}{Y_d^{\rm eq}}\frac{M_P}{M}\left(\frac{T^3-T_i^3}{M^3}\right)\right), \label{Ysol} \ee
 where $Y_i \equiv Y_d(T_i)$ is the initial condition at a given initial temperature $T_i$.
 In Fig.~\ref{YvM} the value of $Y_d$ at the EW scale $T\sim 250$~GeV (above all SM particle masses), which is denoted  $Y_d^{\rm EW}$,   is plotted as a function of $M$ setting $Y_i = 0$.
 By increasing $Y_i$ and/or $T_i$ the value of $Y_d(T)$ increases as clear from~(\ref{Ysol}) because $Y_i$ is always taken to be smaller than $Y_d^{\rm eq}$. Note that, if $Y_d$ was ever zero when the effective Lagrangian~(\ref{Leff}) is valid, the corresponding value of the temperature should be the largest possible one in the range of temperatures where the effective field theory is valid  because the solution in~(\ref{Ysol}) is a decreasing function of $T$ (recall $Y_i<Y_d^{\rm eq}$ and obviously $Y_d(T)\geq 0$).  So~(\ref{Ysol}) and Fig.~\ref{YvM} show that DPs are copiously produced and $Y_d$ can reach its equilibrium value at temperatures above all SM particle masses even if $M$ is several orders of magnitude above the EW scale. Also, note that $Y_d(T)$ decreases exponentially as $T$ increases and so the DP production is most effective at large temperatures, which are as close as possible to the cutoff. This justifies our approximation in which all SM particle masses are neglected in the calculation of the thermal production rate of Sec.~\ref{Thermal production}. 
% Furthermore,  this provides a further reason for avoiding resuming the gauge field propagators
 
 Any given DP model predicts specific coefficients in~(\ref{Leff}) and so a specific value of $\alpha$; one can then compute for any model the DP yield by using Fig.~\ref{YvM} or, more generally, Eq.~(\ref{Ysol}).

  \section{Dark-photon contribution to $N_{\rm eff}^{(\nu)}$}\label{NeffSec}
 
 After being produced, the massless\footnote{For contributions of massive dark photons  to  $N_{\rm eff}^{(\nu)}$ see Refs.~\cite{Jaeckel:2008fi,Ng:2014iqa,Ibe:2019gpv}} DP contributes to the effective number of neutrinos $N_{\rm eff}^{(\nu)}$. Let us determine this contribution here. 
 
%
%Let us start by estimating the decoupling temperature of DPs. 

We  start by computing today's DP temperature. In order to do so one needs to know how many SM degrees of freedom were relativistic at the time of DP decoupling. From the effective Lagrangian in~(\ref{Leff}) one sees that all DP interactions with the SM particles need at least one Higgs field. Therefore,  the effective Lagrangian in~(\ref{Leff}) implies that the DP decoupled  at a (photon) temperature that, at leading order, is not much smaller than the Higgs mass, $T\gtrsim M_h$. Indeed, by including the non-vanishing value of $M_h$ in the calculations of Sec.~\ref{Thermal production} one finds that the thermal production is exponentially suppressed by Boltzmann factors $\sim \exp(-M_h/T)$ for $T<M_h$.  Going to next-to-leading order by switching on SM couplings and/or including effective operators of dimension greater than six, 
one can check  using Eq.~(\ref{BolEq}) that
the DP always decoupled at a temperature not much smaller than the Higgs mass requiring $M$ to be sufficiently above the TeV scale. This requirement ensures that the observational exclusion limits~\cite{Fabbrichesi:2020wbt} are satisfied (recall that the messenger fields have sizable couplings to the SM).
%One can also similarly understand that the DP decoupled at a temperature not much smaller than $M$, the typical mass of the messenger fields, because it is only through the exchange \xxx{what about virtual contributions of the messenger fields?}of those fields that the DP can interact with the SM plasma. 
%When the (photon) temperature was high enough the DP was in thermal equilibrium with all SM particles. We call $T_1$ such a value of the temperature. We take $T_1$ just above the DP decoupling temperature such that all (or almost) all SM particles were relativistic.
  At a temperature $T_1$ just above  the DP decoupling temperature when all 
  (or almost all) 
  SM particles were relativistic 
 the total entropy density of the relativistic plasma  was 
 \be s_1 \simeq (g_{\rm SM}+2)T_1^3\frac{2\pi^2}{45}. \ee 
 After $e^+ e^-$ annihilation the neutrinos decoupled with a temperature $T_\nu = (4/11)^{1/3} T$, where $T$, as always, is the photon temperature. At these later times the entropy density was 
 \be s_2 = \left(2T^3 +\frac78 \times 6 T_\nu^3 + 2 T_d^3\right)\frac{2\pi^2}{45} = \left(\frac{43}{11} T^3 + 2 T_d^3\right)\frac{2\pi^2}{45}, \ee
 where $T_d$ is the DP temperature. From the time when the entropy density was $s_1$ until today the DP was decoupled and, therefore, was not reheated by particle annihilations. As a result $T_d$ scaled at those times as $1/a$, where $a$ is the (cosmological) scale factor. So $T_1 a_1 = T_d a_0$, where $a_1$ and $a_0$ are  the values of the scale factor corresponding  to $T_1$ and to times after $e^+ e^-$ annihilation, respectively. From entropy conservation, $s_1a_1^3=s_2a_0^3$, it follows
 \be T_d = \left(\frac{43}{11 g_{\rm SM}}\right)^{1/3} T. \ee
 
 This DP temperature corresponds today to a contribution to the effective number of neutrinos $N_{\rm eff}^{(\nu)}$ given by 
 \be \Delta N_{\rm eff}^{(\nu)} = \frac87 \left(\frac{T_d}{T_\nu}\right)^4=\frac87 \left(\frac{43}{4g_{\rm SM}}\right)^{4/3}, \label{DeltaNeff}\ee
 which, using $g_{\rm SM}=427/4$, gives
\be \Delta N_{\rm eff}^{(\nu)}  \simeq 0.0535.\ee
 This contribution is twice that of the axion of Quantum Chromodynamics (QCD) determined in~\cite{Salvio:2013iaa} because the DP has two degrees of freedom, but otherwise the DP has properties similar to those of the QCD axion for temperatures above the QCD phase transition\footnote{More recently, the thermal axion production across the QCD phase transition has been computed in~\cite{DEramo:2021psx}.}.
Adding the contribution in~(\ref{DeltaNeff}) to the SM value 3.044 recently computed in~\cite{Bennett:2019ewm,Akita:2020szl,Froustey:2020mcq,Bennett:2020zkv},
 one obtains $N_{\rm eff}^{(\nu)} \simeq 3.1$ in good agreement with the most precise constraint $N_{\rm eff}^{(\nu)}=2.99\pm0.17$
 %3.27\pm0.15
 %
  published by the Planck collaboration in 2018~\cite{Planck:2018vyg}.

 Future measurements of $N_{\rm eff}^{(\nu)}$ will be able to test the DP scenario. For example, the CMB-S4 project (the ``Stage-4" ground-based cosmic microwave background experiment) expects to reach a sensitivity around 0.02-0.03 for $N_{\rm eff}^{(\nu)}$~\cite{CMB-S4:2016ple}.

 \section{Conclusions}\label{Conclusions}
 
 In this paper the thermal production rate of massless DPs has been computed at leading order taking into account the contribution of {\it all} SM particles and using a {\it model-independent} effective field theory point of view. Moreover, the corresponding DP yield and  DP contribution to $N_{\rm eff}^{(\nu)}$ have been calculated too.
 
 In order to perform these computations, in Sec.~\ref{LeffSec} the most general effective field theory describing all possible  interactions with all SM fields up to dimension-six operators has been identified. No dimension-five operators can be constructed describing the interactions of the DP with SM fields only. In all dimension-six operators the Higgs field appears too. 
 
 In Sec.~\ref{Thermal production} the DP thermal production rate per unit of volume, $\gamma_d$, has been computed and the final result has been summarized in Eqs.~(\ref{gammadSum}) and~(\ref{alpha}). The rate grows with temperature as $T^8/M^4$ with a coefficient $\alpha/\pi^3$ that can naturally be as large as$~\sim 10$: this order of magnitude is obtained by assuming all coefficients ($c_i$, $\tilde c_i$, with $i=b,w,p$, and $C_u$, $C_d$ and $C_e$) of the dimension-six operators in~(\ref{Leff}) to be of order 1. Of course, some or many coefficients could be zero in some specific models and $\alpha$ can be smaller in those cases.

The corresponding DP yield $Y_d$ has been computed in Sec.~\ref{Dark-photon yield} solving analytically  the relevant Boltzmann equation. The analytic solution is given in~(\ref{Ysol}) and the value of $Y_d$ at the EW scale is plotted as a function of $M$ in Fig.~\ref{YvM}. These results show that  the DP production is most effective at large temperatures, which are as close as possible to the cutoff compatibly with the validity of the effective field theory, and $Y_d$ can reach its equilibrium value at temperatures above all SM particle masses, even if $M$ is several orders of magnitude above the EW scale. As a result, we also note that the DP production computed here also applies to a massive DP as long as its mass is very small compared to the large temperatures at which the production is maximized. For example, this is always the case when the DP mass is much below the EW scale. For previous computations of massive DP thermal production for temperatures much below the EW scale see Refs.~\cite{Redondo:2008ec,Vogel:2013raa,Ng:2014iqa,Fradette:2014sza}. 

Finally, the corresponding DP contribution to $N_{\rm eff}^{(\nu)}$ has been determined in Sec.~\ref{NeffSec} (see Eq.~(\ref{DeltaNeff})).
%for $g_{\rm SM}=427/4$. 
This prediction is in good agreement with current observations and will be tested with future measurements such as those of CMB-S4.

 \vspace{0.2cm}

 \subsubsection*{Acknowledgments}
 I thank Massimo Bianchi and Marina Migliaccio for useful discussions. 
 This work has been partially supported by the grant DyConn from the University of Rome Tor Vergata.

 \vspace{1cm}
 %\newpage
\footnotesize
\begin{multicols}{2}

\end{multicols}

\end{document}